\documentclass[apjl]{emulateapj}

\usepackage{amssymb}
\usepackage{epsfig}
\usepackage{longtable} 
\usepackage{amsmath}
\usepackage{amsfonts}
\usepackage{ctable}
\usepackage{graphics}
\usepackage{subfigure}
\usepackage{verbatim}
\usepackage{hyperref}
\newcommand{\lp}{\left (}
\newcommand{\rp}{\right )}

\newcommand{\ra}{\rightarrow}

\newcommand{\p}[2]{\frac{\partial #1}{\partial #2}}

\newcommand{\ev}[1]{\left \langle #1 \right \rangle}

\def\mnras{MNRAS}
\def\apjl{ApJl}

\def\icarus{Icarus}
\begin{document}

\title{Statistical Signatures of Panspermia in Exoplanet Surveys}

\author{Henry W. Lin\altaffilmark{1,2}, Abraham Loeb\altaffilmark{2}}
%\author{Henry W. Lin$^{1}$\thanks{E-mail: henrylin@college.harvard.edu; ggonzalezabad@cfa.harvard.edu; aloeb@cfa.harvard.edu}, 
%Gonzalo Gonzalez Abad$^{2}$\footnotemark[1]
%and Abraham Loeb$^{2}$\footnotemark[1]\\
%$^{1}$
\altaffiltext{1}{Harvard College, Cambridge, MA 02138, USA}
\altaffiltext{2}{Harvard-Smithsonian Center for Astrophysics, 60 Garden St., Cambridge, MA 02138, USA}
\altaffiltext{}{Email: henry.lin@cfa.harvard.edu, aloeb@cfa.harvard.edu}

%}$^{2}$Harvard-Smithsonian Center for Astrophysics, 60 Garden St., Cambridge, MA 02138, USA}

%\date{Accepted 1988 December 15. Received 1988 December 14; in original form 1988 October 11}
%\date{Accepted MMMM. Received NNNN; in original form 2014 June 11}
%\pagerange{\pageref{firstpage}--\pageref{lastpage}} \pubyear{2014}

%\label{firstpage}

\begin{abstract}
A fundamental astrobiological question is whether life can be transported between extrasolar systems. We propose a new strategy to answer this question based on the principle that life which arose via spreading will exhibit more clustering than life which arose spontaneously. We develop simple statistical models of panspermia to illustrate observable consequences of these excess correlations. Future searches for biosignatures in the atmospheres of exoplanets could test these predictions: a smoking gun signature of panspermia would be the detection of large regions in the Milky Way where life saturates its environment interspersed with voids where life is very uncommon. In a favorable scenario, detection of as few as $\sim 25$ biologically active exoplanets could yield a $5\sigma$ detection of panspermia. Detectability of position-space correlations is possible unless the timescale for life to become observable once seeded is longer than the timescale for stars to redistribute in the Milky Way.

%We revisit the Fermi paradox and comment on the relevance of the equilibrium properties of these models to the future fate of life in the universe.

\end{abstract}

\maketitle

%=================
\section{Introduction}
The question of where life originated is centuries old \cite[for a review, see][]{miller74,review}, but to date the only experimentally viable method of detecting panspermia is the detection of biomaterial on an asteroid or comet. Unless a significant fraction of interplanetary objects are biologically active, this method will not yield positive results or falsify the hypotheses of panspermia because the enormous number of objects in our solar system \citep{moro} may permit a significant number of panspermia events, even if the fraction of objects which contain life is miniscule. Although previous estimates suggested that lithopanspermia events should be quite rare \citep{melosh, lith}, more recent proposals \cite{belbruno} yield considerably more optimistic rates. Given the experimental difficulties of testing the hypotheses of panspermia and poor constraints on the theoretical diversity of life, one may even question whether panspermia is truly falsifiable. In this {\it Letter}, we answer this question in the affirmative. Under certain conditions, panspermia leads to statistical correlations in the distribution of life in the Milky Way. If future surveys detect biosignatures in the atmospheres of exoplanets, it will be possible to devise statistical tests to detect or constrain panspermia event rates while remaining agnostic to the biological mechanisms of panspermia.

This {\it Letter} is organized as follows. In \S 2 we describe a simple class of panspermia models that qualitatively captures the statistical features of any panspermia theory. We discuss observable signatures of panspermia in \S 3. We conclude with further implications of our panspermia models in \S 4.
\begin{figure*}
	\centering
	\subfigure{\includegraphics[height=0.2\textheight, clip=true]{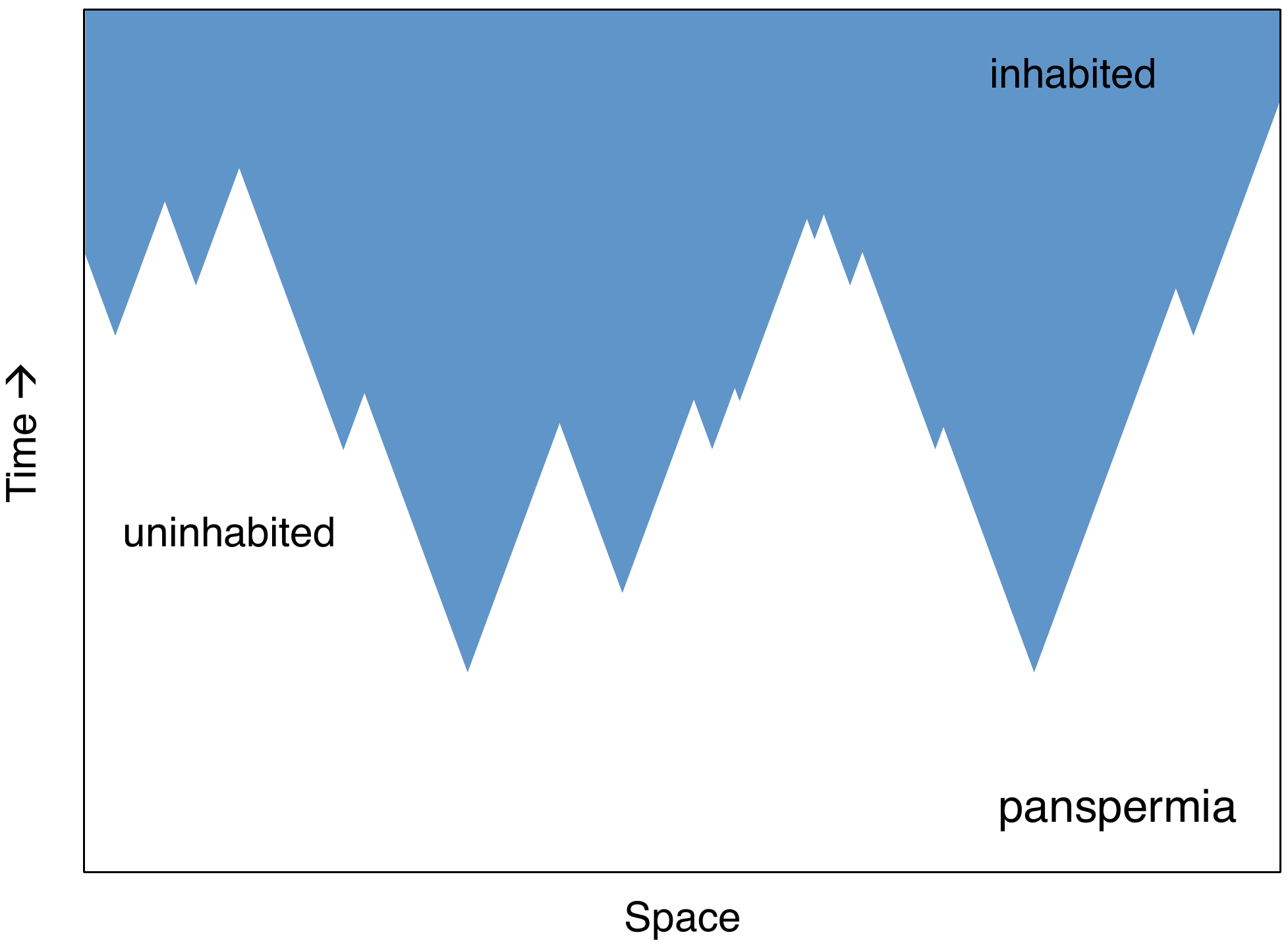}}
	\hspace{1cm} 	
	\subfigure{\includegraphics[height=0.2\textheight, clip=true]{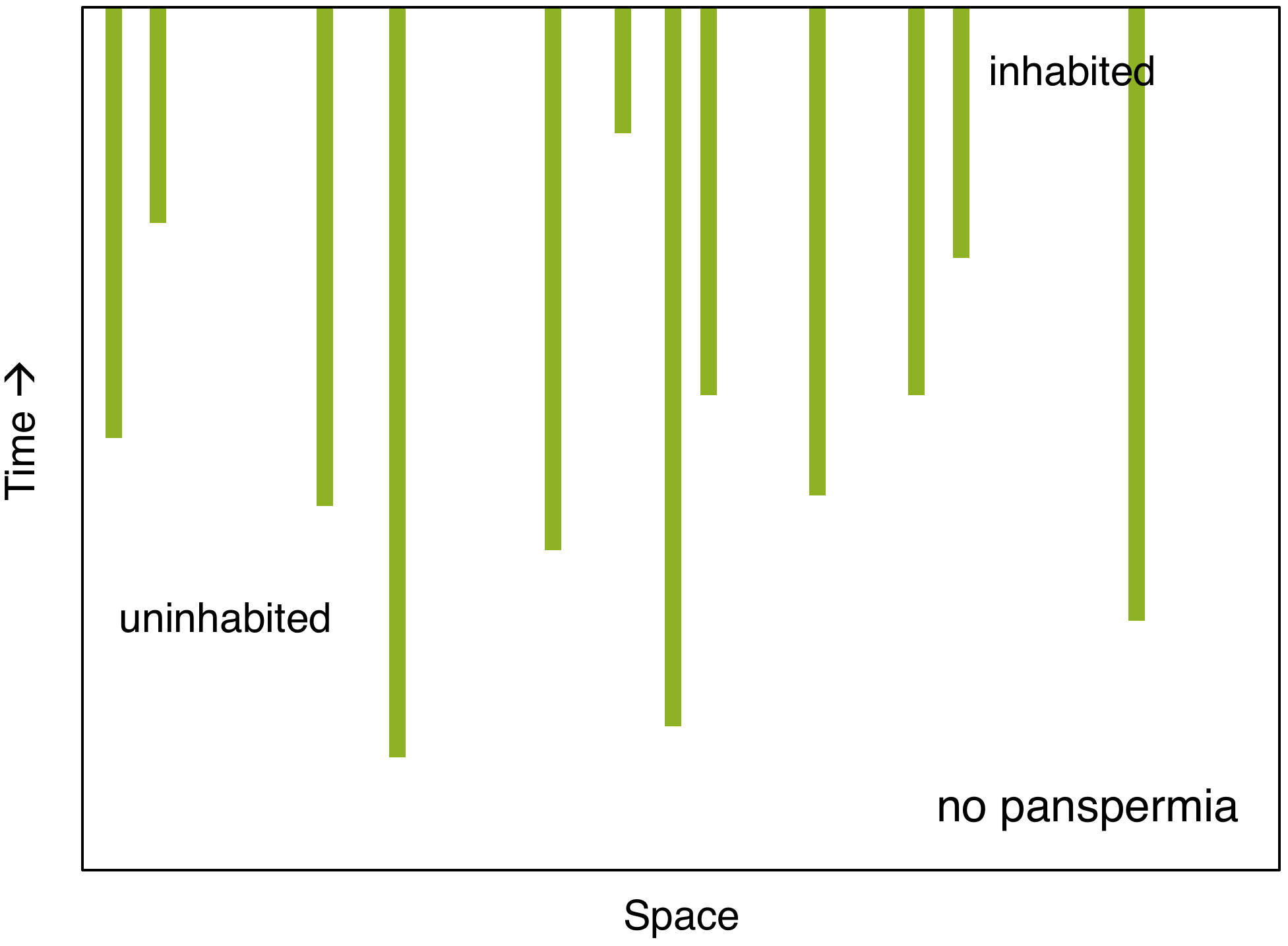}}
 	\caption{
	Schematic diagrams of the topology of the bio-inhabited planets within the galaxy for the panspermia case (left) and no panspermia case (right). In the panspermia case, once life appears it begins to percolate, forming a cluster that grows with time. Life can ocassionally spontaneously arise after the first bio-event, forming clusters that are smaller than more mature clusters. (The limiting case where life spontaneously arises once and then spreads to the rest of the galaxy would correspond to a single blue triangle. In the ''sudden'' scenario, all triangles start at the same cosmic time and are thus the same size.) As time progresses, the clusters eventually overlap and the galaxy's end state is dominated by life. In the no panspermia scenario, life cannot spread: there is no phase transition, but a very gradual saturation of all habitable planets with life. Observations of nearby habitable exoplanets could statistically determine whether panspemia is highly efficient (left), inefficient (right), or in some intermediate regime. \label{comp}
	}
\end{figure*}

\section{A Model for Panspermia}
%Any dynamical system involves the specification of initial conditions and rules for updating the system with time. Our model is no exception. 

Consider an arbitrary lattice $L$ in two or three dimensions. (The two dimensional model corresponds to a thin-disk approximation of the Milky Way). While a lattice model is a crude approximation to reality, lattices are analytically tractable, and the conclusions we will draw will hold in the continuum limit.\footnote{A slightly more realistic model would distribute the points randomly and consider circles or spheres of influence surrounding these points where panspermia events could take place. This setup is known as a continuum percolation problem in the mathematics literature \citep{meester}. In the regime where most of the spheres of influence overlap, the lattice approximation described above gives similar results.} Each lattice point represents a habitable extrasolar system. Viewed as a graph, the number of edges $N$ associated with each lattice point represents the average number of panspermia events per extrasolar system. Associated with each point ${\bf x}\in L$ is a state variable $h({\bf x})$ which is either 0 or 1, representing the biologically uninhabited and inhabited states, respectively. 

The initial state of the lattice is $h({\bf x}) = 0$ for all ${\bf x} \in L$. The system is updated as follows. At each discrete time step, neighbors of each inhabited site become inhabited. Furthermore, some fraction $0 < p_s < 1$ of the uninhabited sites are switched to $h=1$. This describes panspermia in the regime where life spontaneously arises at a very gradual constant rate. It is also possible to study the opposite regime, where life spontaneously arises suddenly. We will refer to the two regimes as the ``adiabatic'' and ``sudden'' scenarios. 

We consider the adiabatic case first. Consider the regime where $p_s \ll 1$ and $N \ge 1$. Pictorially, bubbles form in the lattice as shown in Figure 1. At each time step, the bubbles grow linearly in size, as new bubbles are formed. After a while, there are bubbles of many sizes. Well before the overlap time, namely for $t \ll t_o$, the probability that a lattice site is contained in a bubble of radius $R$ is given by

\begin{equation}
p(R) \approx p_0(1- V(R+1) p(R+1)),
\end{equation}
where $R_\text{max} = t$ and $p(R_\text{max}) = p_0$ and $V(R)$ in two (three) dimensions is the area (volume) of a bubble of radius $R$.
Since bubbles of size $R+1$ already occupy some of the lattice sites, new, smaller bubbles will have less room to form, which is reflected in the second term in equation (1). However, to linear order in $p_0$, the distribution is uniform on the interval $[0, R_\text{max}]$. In the regime where bubbles of all sizes are present with approximately equal number densities, the system resembles a fluid at a scale-invariant critical point \citep{stanley}, where bubble nucleation converts one phase to the other. This scenario is analogous to the formation of HII bubbles during the epoch of reionization \citep{loeb13} or the production of bubbles during cosmic inflation \citep{turner}.

In the sudden case, the initial conditions are such that each site is inhabited with probability $p_s$. The dynamical rule for updating the system is simply that neighbors adjacent to an inhabited site become inhabited. In this case, all bubbles are formed with the same size, and grow linearly as a function of time. The dynamics of the system can also be easily described via renormalization group methods. Consider the operation of partitioning the lattice into blocks containing a fixed number $\ell$ of lattice sites. The renormalized probability $p'$, e.g. the probability that at least one of $\ell$ sites will be habitable is $p'=1 - (1 - p_s)^\ell$. Note that $p' > p_s$ for $0 < p_s < 1$ and $p'$ is an increasing function of $\ell$. The coarse grained evolution of the system corresponds to incrementing $\ell$ with each time step and setting all sites in a block of size $\ell$ to $h=1$ if a single site is inhabited. The dynamical picture is one where clusters of life form and grow, overlap, and eventually merge into a percolating cluster when the renormalized probability $p'$ equals the percolation threshold for the given lattice.

A qualitative difference between a simple lattice model and the Milky Way is that stars in the Milky Way drift relative to each other with a characteristic speed $\sigma_v$ of a few tens of km/s \citep{binney2011}. This presents three interesting regimes which are characterized by the effective spreading speed of life $v$.  The effects of drifting should be negligible in the limit that panspermia takes place at speeds $v \gg \sigma_v$. This could be the case if there exists an intelligent species which can spread at high speeds. But even if panspermia takes place at speeds comparable to the relative speeds of stars, the results from our lattice models still hold. To see this, consider modifying the dynamical rule in the following way: at each given time increment, a cell is randomly swapped with one of its nearest neighbors. This simulates the random motion of stars due to their velocity dispersion. Consider a bubble which has already formed. Outside of the bubble, swapping uninhabited lattice sites has no effect on the correlation function. Inside the bubble, swapping uninhabited lattice sites also does not produce any observable effects. Only the boundary of the bubble is affected by swapping. If the dimensionality of the lattice is large, most of the lattice sites on the boundary will be swapped amongst themselves. In a two or three dimensional lattice, the boundary effectively grows by $< 1$ unit. Macroscopically, the effective speed of panspermia increases by a factor of order unity. This regime may be of particular interest to lithopanspermia, since ejected rocks have velocities $v \sim \sigma_v$. Finally, the third regime is when panspermia takes place at a rate $v \ll \sigma_v$. Although a proper treatment of this regime requires numerically integrating orbits of stars in the Milky Way, a tractable approximation is given by a linearized reaction-diffusion equation, which describes randomly-walking stars that can spread life locally:
\begin{equation}
\p{h}{t}= D \nabla^2 h + \Gamma h + J,
\end{equation}
where $D$ is the diffusion constant that controls the relative drifting of stars, $\Gamma$ is the infection rate, and a source term $J$ accounts for the spontaneous development of life. In the adiabatic regime, $J$ is the sum of delta functions uniformly distributed on some region of space-time. In the sudden regime, $J$ consists of a single delta function. To determine the evolution of this system, it suffices to compute the Green's function $G(\textbf{x},\textbf{x}_0)$ for this equation:
\begin{equation}
G(\textbf{x},\textbf{x}_0) = e^{\Gamma t} \times \Delta\lp\frac{\textbf{x}-\textbf{x}_0}{\sqrt{2D t}}\rp,
\end{equation}
where $\Delta$ is a Gaussian with vanishing mean and a standard deviation of unity. Note that the $\Delta$ term is simply the propagator for the diffusion equation. In this regime, bubble formation is modified by the fact that bubbles grow in size $R \propto (Dt)^{1/2}$ instead of $R \propto t$. 

It should be noted that the results stated above will hold until bubbles are too large to neglect the effects of velocity shear in the Milky Way \citep{binney2011}. Once the bubbles grow to a significant fraction the radius of the Milky Way, they will be sheared apart on $\sim 100$ Myr timescales. Hence, shearing effectively disperses bubbles greater than some critical size. If the density of the dispersed region is low, the separation between inhabited sites will be large, so the shearing will effectively convert convert the larger bubble into smaller bubbles. So long as the phase transition is not complete, small bubbles will start to regrow, and the cycle starts over again.

Finally, it is important to note that life may take a non-negligible amount of time $t_d$ to become detectable once life is spread to it. We will not attempt to quantify $t_d$ except to note that $t_d$ could in principle be very short if a photosynthetic (or more exotically, an industrially polluting) species can propagate between solar systems such as cyanobacteria \citep{cyano}. Of course, for Earth's history $t_d$ (for currently proposed biomarkers) is very long $t_d \sim 10^{9}$ yr. If $t_d$ is much larger than the timescale for stars to diffuse/shear (on the order of $\sim 10^8$ yr), it will not be possible to detect any position-space correlations, though more subtle phase-space correlations (see \S 3) could in principle be detected.

\begin{figure}
	\centering
	\includegraphics[height=0.2\textheight, clip=true]{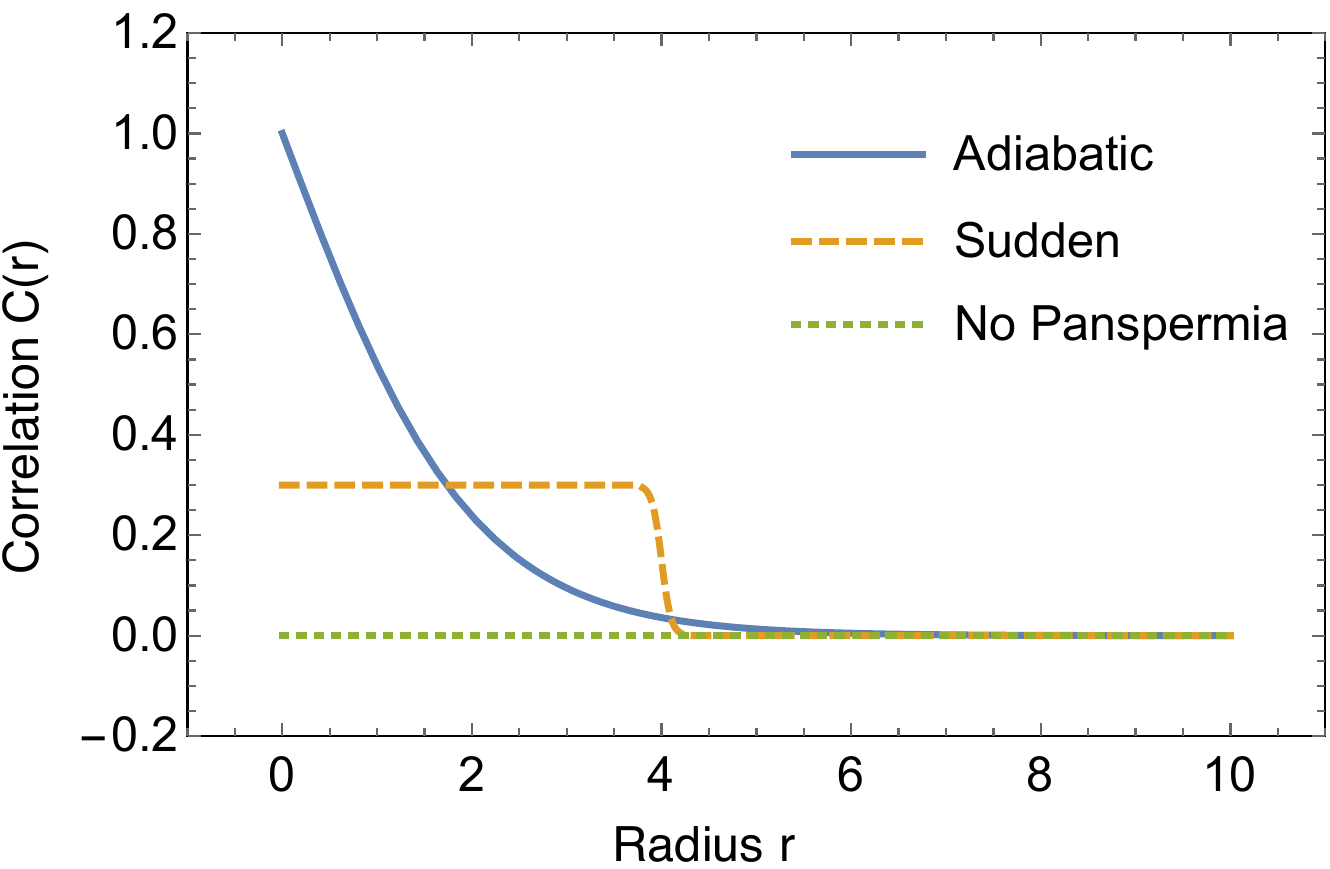}
 	\caption{
	Schematic plot of the spherically averaged correlation function $C(r)$ as a function of radius $r$  during the panspermia phase transition for two different panspermia scenarios and the no panspermia scenario. In the adiabatic regime, life can percolate in addition to spontaneously arising. In the sudden regime, life arises once and then begins to percolate. Both correlation functions define a characteristic scale radius $\xi$ which reflects the time elapsed $t$  since the first life started percolating. In the case where no panspermia occurs, the reduced correlation function is exactly zero, so a measurement of $\mathcal{C}(r) > 0$ (see equation (5) for a definition) would be compelling evidence for panspermia. 
	\label{comp}
	}
\end{figure}

\section{Observable signatures}
An important observable consequence of panspermia that is illustrated in both of these models. The two-point correlation function
\begin{equation}
C(\textbf{x}-\textbf{y}) \equiv \ev{h(\textbf{x})h(\textbf{y})} - \ev{h(\textbf{x})}^2
\end{equation}
has the property $C \ne 0$ during the entire evolution of the system. This is true unless we are unlucky enough that the phase transition has already been completed and $h({\bf x})=1$ everywhere. In particular, the timescale for the phase transition to run to completion is most likely several times the life crossing time of the Milky Way. For $v \sim 10$ km/s, the life crossing time corresponds to several Gyr, so the phase transition could take of order the Hubble time. For the sudden case, the correlation length $\xi \sim \text{min}(\text{max}(v t, \sqrt{2D t}), R_\text{MW})$ where $R_\text{MW}$ is a length scale several times smaller than the radius of the Milky Way. For the adiabatic case, the correlation function is peaked at $\textbf{x}-\textbf{y} = 0$ and drops to zero over the same characteristic length $\xi$. In general, $\xi$ will always show a cutoff at some scale radius $\sim R_\text{MW}$ due to shearing effects. The schematic form of the correlation function is displayed in Figure 2. A more complicated rule where the rate $p_s$ varies with time will encode itself in the correlation function; the important point is that any spreading whatsoever will yield potentially observable deviations from $C =0$ which is the Poisson case.

For real observations, one must take into account the fact that stars are not distributed uniformly on a lattice and may themselves exhibit clustering. To take this into account, we propose the following {\it reduced} correlation function as a potentially robust indicator of panspermia:

\begin{equation}
\mathcal{C}(\textbf{x}-\textbf{y}) \equiv \ev{\frac{h(\textbf{x})h(\textbf{y})}{\sigma(\textbf{x})\sigma(\textbf{y})}}-\ev{\frac{h(\textbf{x})}{\sigma(\textbf{x})}}^2,
\end{equation}
where in the continuum limit $\sigma(\textbf{x})$ is the stellar density and $h(\textbf{x})$ is the density of inhabited stellar systems. If life arises independently among different stellar systems, $h \propto \sigma$, so $\mathcal{C}(\textbf{x}-\textbf{y})$ accounts for the fact that stars do not form a perfect lattice by ``dividing out'' the star-star correlation. An even more sophisticated treatment could replace the spatial densities with corresponding phase space densities $\textbf{x}\ra\lp \textbf{x},\textbf{p}\rp$, since two stars which are closer together in phase space will have more time to transfer biomaterial than two stars which are close in position space but far away in momentum space. In principle, one could reverse-integrate the orbits of stars and calculate the radius of closest approach $r_c$ for any two given stars. Measuring the correlation as a function of $r_c$ would be an alternative strategy for disentangling the effects of stellar diffusion and panspermia, which may be useful if life propagates at very low speeds, or if the timescale $t_d$ is longer than the stellar mixing timescale. %In the continuum approximation, Liouville's theorem guarantees that the phase space volume of bio-inhabited stars grows only due to panspermia events. However, close stellar encounters (not to mention imperfections in our knowledge of the galactic potential) will increase the effective phase space volume. Precise knowledge of the distribution of inhabited stars in phase space would determine in principle the location of the stars at any time in the past. In practice, this may not be practical for all but the nearest stars. 

We note that the above discrete models can be generalized to the continuous case by using a slightly different formalism. If the number distribution of bubbles is known, a power spectrum of inhabited star density fluctuations can be derived that will reproduce the bubble spectrum by retracing the steps of the Press-Schechter formalism \citep{ps74}. Once the power spectrum is obtained, one can obtain the correlation function via a Fourier transform. If the habitable stellar density is not constant but fluctuates in space, the problem becomes analogous to the spread of disease on an inhomogeneous medium \citep{lin15}, which again makes use of the Press-Schechter formalism.

Future surveys such as the TESS \citep{tess} will detect hundreds of earth-like exoplanets \citep{tess2}. Ground-based and space-based (e.g. JWST) follow ups that can characterize the exoplanet atmospheres could test for biosignatures such as oxygen in combination with a reducing gas \citep[for a review, see][]{kalt}. However, it is likely that only a few earth-like exoplanets will be close enough to be biologically characterized \citep{brandt, rein14} with next-decade instruments. Eventually, surveys could test for more specific spectral signatures such as the ``red edge'' of chlorophyl \citep{rededge} or even industrial pollution \citep{lin}. It is also possible that searches for extraterrestrial intelligence in the radio or optical wavelengths could also yield detections that could be tested for clustering. Any positive detections will yield first constraints on the correlation function $C(\textbf{x}-\textbf{y})$. In a favorable scenario, our solar system could be on the edge of a bubble, in which case a survey of nearby stars would reveal that $\sim 1/2$ of the sky is inhabited while the other half is uninhabited. In this favorable scenario, $\sim 25$ targets confirmed to have biosignatures (supplemented with 25 null detections) would correspond to a $5 \sigma$ deviation from the Poisson case, and would constitute a smoking gun detection of panspermia. A more generic placement would increase the number of required detections by a factor of a few, though an unusual bubble configuration could potentially reduce the number of required detections. It should be noted that the local environment of our solar system does not reflect the local environment $\sim 4$ Gyr ago when life arose on earth, so the discovery of a bubble of surrounding earth should be interpreted as the solar system ``drifting'' into a bubble which has already formed, or perhaps the earth seeding its environment with life.
%If our solar system is inside a large inhabited bubble, this would provide convincing evidence that life was brought to earth from another stellar system, since the probability that the earth was the ``seed'' of the bubble decreases with the number of sites contained in the bubble. The odds that panspermia initiated life on earth would be further augmented if the earth was near the edge of the bubble, since the seed of the bubble is much more likely to be near the center.

Finally, we note that since the efficiency of panspermia is dictated by the number of transfer events between stellar systems, which in turn will depend on the local stellar density. For example, the transfer rates of rocky material between stars grows with stellar density simply because the mean distance between stars is smaller \citep{lith, belbruno}, and the enhanced frequency of close stellar encounters should further increase these rates. One might therefore expect that an inhomogeneous distribution of stars (i.e. the Milky Way) might exhibit a ``multi-phase'' structure, where regions of high stellar density have completed the phase transition and are saturated with life, whereas regions of low stellar density will have little to no signs of life. However, regions of high stellar density may be inhospitable to life due to a variety of factors \citep{ghz, ghz2, ghz3}. For example, perturbations of Oort cloud analogs due to close stellar encounters, could potentially make a habitable zone planet inhospitable to life \citep{ghz}. The number density per unit time of supernovae explosions \citep{sn} and the local stellar metallicity may also play a factor in determining the habitability of planets \citep{ghz}. It is therefore likely that the number of panspermia events is a complicated, non-monotonic function of stellar density and other astrophysical parameters.

\section{Implications}
In our simple formalism, life spreads from host to host in a way that resembles the outbreak of an epidemic. A key point is that the correlations that quickly arise imply that the transition from an uninhabited galaxy to an inhabited galaxy can occur much faster in the panspermia regime than in the Poisson case. Panspermia implies a phase transition, whereas a Poisson process will only lead to a gradual build up of life. Said differently, the start time for life for different stellar systems exhibits a very small scatter in the panspermia case. A consequence of the panspermia scenario is that the severity of the Fermi paradox may be reduced somewhat. If life started everywhere at the same time, we expect fewer advanced civilizations at the present time than if life could have started much earlier on other stellar systems. It should be noted, however, that this statement is predicated on the somewhat controversial assumption that there is an evolutionary bias towards increasing complexity \citep{adami}. A second consequence of panspermia is that the Drake equation \citep{shk66} becomes a lower bound on the number of civilizations, since the multiplicative form of the equation is based on the assumption that life arises independently everywhere. This assumption may be strongly violated in the regime where panspermia is highly efficient.

The mathematical similarity between panspermia and disease spread may also represent a biological one: any species which acquires panspermia abilities will have enormous fitness advantages. Just as viruses evolved to brave the ``harsh'' environment of ``inter-host'' space to harness the energy of multiple biological hosts, perhaps evolution has or will drive a class of organisms to brave the harsh environment of interstellar space to harness the energy of multiple stellar hosts. Whether or not the organisms will be primitive (e.g., lithopanspermia, cometary panspermia \citep{hoyle}) or intelligent (directed panspermia \citep{crick} or accidental panspermia) remains to be determined.

Even if the earth is the only inhabited planet and primitive life cannot survive an interstellar journey, interstellar travel led by humans may one day lead to colonization of the galaxy. As a zeroth order model, the same formalism should approximately describe the growth of the colonies as they percolate through the galaxy, assuming that the processes such as population diffusion also occur in space \citep{newman}. Although the question is a purely astrobiological question today, in the distant future urban sociologists and astrophysicists might be forced to work together \citep{lin15}. Indeed, well after the colonization regime, models of panspermia may continue to be relevant, as spaceships capable of interstellar travel will provide the opportunity for primitive life (e.g. domesticated life, diseases, and viruses) to spread efficiently. The question that awaits is whether primitive life has already spread efficiently, or whether it will have to wait for ``intelligent'' life to make the voyage.

\section*{Acknowledgments}
We thank Ed Turner, Edwin Kite, and an anonymous referee for useful discussions. This work was supported in part by NSF grant AST-1312034, and the Harvard Origins of Life Initiative.

%\bibliographystyle{apj}
%\bibliography{panspermia}{}

\end{document}